\documentclass[pra,eqsecnum,showpacs,twocolumn]{revtex4}
\usepackage{graphicx}
\usepackage[urlcolor=blue, hyperindex, colorlinks, bookmarks=true,linkcolor=black,citecolor=black]{hyperref}

\begin{document}
\title{Dynamical parameter estimation using realistic photodetection}
\author{P. Warszawski}
\affiliation{Centre for Quantum Computer Technology,
Centre for Quantum Dynamics, School of Science, Griffith
University,  Brisbane 4111, Australia}
\author{Jay Gambetta}
\affiliation{Centre for Quantum Computer Technology, Centre for
Quantum Dynamics, School of Science, Griffith University,
Brisbane 4111, Australia}
\author{H. M. Wiseman}
\email{h.wiseman@griffith.edu.au} \affiliation{Centre for Quantum
Computer Technology, Centre for Quantum Dynamics, School of
Science, Griffith University, Brisbane 4111, Australia}
\newcommand{\beq}{\begin{equation}}
\newcommand{\eeq}{\end{equation}}
\newcommand{\bqa}{\begin{eqnarray}}
\newcommand{\eqa}{\end{eqnarray}}
\newcommand{\nn}{\nonumber}
\newcommand{\nl}[1]{\nn \\ && {#1}\,}
\newcommand{\erf}[1]{Eq.~(\ref{#1})}
\newcommand{\erfs}[2]{Eqs.~(\ref{#1})--(\ref{#2})}
\newcommand{\dg}{^\dagger}
\newcommand{\rt}[1]{\sqrt{#1}\,}
\newcommand{\smallfrac}[2]{\mbox{$\frac{#1}{#2}$}}
\newcommand{\half}{\smallfrac{1}{2}}
\newcommand{\bra}[1]{\langle{#1}|}
\newcommand{\ket}[1]{|{#1}\rangle}
\newcommand{\ip}[2]{\langle{#1}|{#2}\rangle}
\newcommand{\sch}{Schr\"odinger }
\newcommand{\schs}{Schr\"odinger's }
\newcommand{\hei}{Heisenberg }
\newcommand{\heis}{Heisenberg's }
\newcommand{\bl}{{\bigl(}}
\newcommand{\br}{{\bigr)}}
\newcommand{\ito}{It\^o }
\newcommand{\str}{Stratonovich }
\newcommand{\dbd}[1]{{\partial}/{\partial {#1}}}
\newcommand{\sq}[1]{\left[ {#1} \right]}
\newcommand{\cu}[1]{\left\{ {#1} \right\}}
\newcommand{\ro}[1]{\left( {#1} \right)}
\newcommand{\an}[1]{\left\langle{#1}\right\rangle}
\newcommand{\st}[1]{\left|{#1}\right|}
\newcommand{\implies}{\Longrightarrow}
\newcommand{\tr}[1]{{\rm Tr}\sq{ {#1} }}
\newcommand{\del}{\nabla}
\newcommand{\du}{\partial}
\newcommand{\singlecol}{\end{multicols}
     \vspace{-0.5cm}\noindent\rule{0.5\textwidth}{0.4pt}\rule{0.4pt}
     {\baselineskip}\widetext }
\newcommand{\doublecol}{\noindent\hspace{0.5\textwidth}
     \rule{0.4pt}{\baselineskip}\rule[\baselineskip]
     {0.5\textwidth}{0.4pt}\vspace{-0.5cm}\begin{multicols}{2}\noindent}
\newcommand{\tick}{$\sqrt{\phantom{I_{I}\hspace{-2.1ex}}}$}
\newcommand{\cross}{$\times$}
\newcommand{\ww}{which-way }%{{\em welcher Weg }}
\newcommand{\ps}[1]{\hspace{-5ex}{\phantom{\an{X_{w}}}}_{#1}\!}
\newcommand{\bbb}[1]{\hspace{-5ex}{\phantom{\an{X_{w}}}}_{#1}\!}
\newcommand{\mket}[1]{|\hspace{-0.3ex}\ket{#1}\hspace{-0.5ex}\rangle}
\newcommand{\mbra}[1]{\langle\hspace{-0.5ex}\bra{#1}\hspace{-0.3ex}|}

\newcommand{\red}[1]{\color{red}{#1}\color{black}}
\newcommand{\blu}[1]{\color{blue}{#1}\color{black}}

\newcommand{\raro}{\rightarrow}
\newcommand{\ve}{\varepsilon}
\newcommand{\bfi}{{\bf I}_{[0,t)}}
\newcommand{\omtrue}{\Omega_{{\rm true}}}
\newcommand{\ommax}{\Omega_{{\rm max}}}
\newcommand{\gamr}{\gamma_{{\rm r}}}
\newcommand{\dtime}{\tau_{{\rm dd}}}
\newcommand{\dwj}{d{\cal W}_{{\rm J}}(t)}
\newcommand{\dwjp}{dW_{{\rm J}}(t)}

\begin{abstract}
We investigate the effect of imperfections in realistic detectors upon the problem of quantum state and parameter estimation by continuous monitoring of an open quantum system. Specifically, we have reexamined the system of a two-level atom with an unknown Rabi frequency introduced by Gambetta and  Wiseman [Phys. Rev. A {\bf 64}, 042105 (2001)]. We consider only direct photodetection and use the realistic quantum trajectory theory reported by Warszawski, Wiseman, and Mabuchi [Phys. Rev. A {\bf 65}, 023802 (2002)]. The most significant effect comes from a finite bandwidth, corresponding to an uncertainty in the response time of the photodiode. Unless  the bandwidth is significantly greater than the Rabi frequency, the observer's ability to obtain information about the unknown Rabi frequency, and about the state of the atom, is severely compromised. This result has implications for quantum control in the presence of unknown parameters for realistic detectors, and even for ideal detectors, as it implies that most of the information in the measurement record is contained in the precise timing of the detections.
\end{abstract}


\maketitle

\section{Introduction}

Quantum parameter estimation
originated with the work by Helevo \cite{Hol82} and
Helstrom \cite{Hel76} and is usually formulated as follows.
A known quantum state undergoes some
evolution. This evolution is usually unitary but need not be
\cite{ChuNie97}, and is parameterized by one or more unknown
parameters. The goal is to estimate these parameters by making a
measurement on the final state. In a recent paper, two of us
\cite{Jay}, following Mabuchi \cite{MabuchiEst}, considered the
situation in which an open quantum system parameterized by an
unknown dynamical parameter is estimated through
continuous-in-time measurements of the environment. Ref.
\cite{Jay} also focussed on the conditioned state and how the lack
of information about the parameter affects its purity. These two
papers are part of a small but growing body of work including
\cite{DohertyStateDeterm} and \cite{DohertyOptQPE} (and references
therein) that address the problem of reducing {\em classical}
uncertainties in the evolution of quantum systems by performing
continuous-in-time measurements.

In Ref.~\cite{Jay}, the system is taken to be a damped,
classically driven two-level atom (TLA) with a driving Hamiltonian
($\propto\hat\sigma_{x}$ in a rotating frame) with an uncertain
driving strength $\Omega$.  The physical scenario the authors give
for this is as follows.  An atom is placed in a classical standing
wave, thus experiencing a Rabi frequency of \beq
\Omega=\Omega_{{\rm max}}\sin(kx), \label{drive} \eeq where $k$ is
the wavevector for the classical field and $x$ is the position of
centre of mass of the atom.  Gambetta and Wiseman then assume that
the TLA is equally likely to be placed at any position in the
standing wave.  Hence, a range of $-\Omega_{{\rm
max}}\leq\Omega\leq\Omega_{{\rm max}}$ exists for the possible
driving strength.  It is also assumed that the TLA has zero
translational motion.  This can be justified by assuming that the
TLA is fixed by a radio frequency Paul trap or similar confining
mechanism \cite{MabuchiEst}.  Although it is shown that continual
efficient measurement upon the output electromagnetic field of a
TLA leads to information gain about the driving strength, the
amount of information gained depends strongly on the measurement
scheme used.  The purity of the conditional state also depends
strongly on the measurement scheme, but the performance of schemes
by these different criteria are {\em not} well correlated at all
\cite{Jay}.

In this paper, we consider only one sort of detection: direct
detection, but we consider {\em imperfect} detection. This allows us to address
the following two  questions. First, what (qualitatively) are the
 aspects of the ideal measurement record that contain the information about
 the system state and the unknown parameter?
 Second, in practice, how badly would the amount of information gained
 be affected by detectors with realistic imperfections.
  To treat realistic detection we
use the theory developed in
Refs.~\cite{WarWisMab02,WarWis03a,WarWis03b}. This generalized the
theory of quantum trajectories
\cite{Car93b,BelSta92,Bar93,DalCasMol92,GarParZol92,Wis96a} by
determining the state of the system conditioned upon the output of
a detector that has dead-time ($\tau_{\rm dd}$), dark counts at
rate $\gamma_{\rm dk}$, inefficiency $\eta$, and finite bandwidth
$\gamma_{\rm r}$. A diagram of this detector is displayed in Fig.
\ref{Model}.

\begin{figure}
\includegraphics[width=0.45\textwidth]{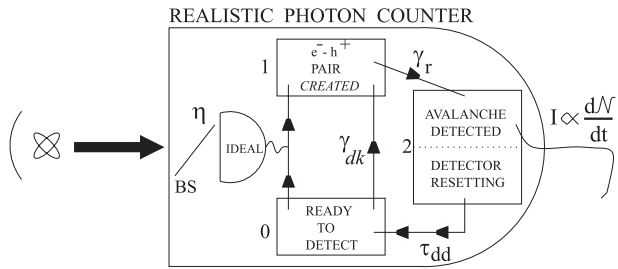}
\vspace{0.2cm} \caption{A schematic illustrating realistic direct
detection. The atom is placed at the focus of a parabolic mirror
such that all the fluorescence enters the realistic detector.
Realistic photon counting is performed by an avalanche photodiode.
The quantum efficiency $\eta$ of the diode is represented by the
the beam-splitter (BS). Single arrowheads within the realistic
photodetector indicate Poisson processes. The microscopic detector
states are indicated by a 0, 1, and 2. } \protect\label{Model}
\end{figure}

The reciprocal of the bandwidth $\gamma_{\rm r}^{-1}$ sets the
scale for the uncertainty in the delay between the absorption of
photon (which is synonymous with its detection in an ideal
detector) and the avalanche of charge which registers this at a
macroscopic level in a realistic detector. We find that a finite
bandwidth has the greatest effect on our parameter estimation
problem. This tells us that  it is the timing between detections, not the steady
state detection rate, that contains the majority of the information gained.
Unless the bandwidth of the detector is
substantially greater than $\Omega_{\rm max}$, the rate of gain of
information about $\Omega$ is much reduced, and consequently the
purity of the system is also. Like the work of Ref.~\cite{Jay}
this has important implications for the robustness of quantum
feedback control (see for example
Refs.~\cite{Dohe2000,WisManWan02}).

This paper is structured as follows. In Sec.~II we begin with some
numerical results to motivate the theoretical developments in the
remainder of the paper. In Sec.~III we briefly review the theory
of parameter estimation based on linear quantum trajectories in
Ref.~\cite{Jay}, and show how it can be applied to realistic
detection. In Sec.~IV we use this theory to investigate the
evolution of the conditional state for an initially unknown
parameter, and in Sec.~V we investigate the rate of information
gain about the parameter. We conclude with a discussion in
Sec.~VI.

\section{Conditional Probability Distribution for Driving Strength}
\label{condprob}
Before providing a theoretical basis,
some numerical results are provided to give the reader an
insight into the effects of realistic detection on parameter
estimation.  With an example in mind, the theory contained in the
following section should also
prove to be more transparent.

We begin by looking at how an observer's probability distribution
for the driving strength evolves given a single measurement record
(trajectory).  Details of the numerical techniques employed to do
this will be given at the end of the next section.  The
distribution is denoted by $P(\Omega|\bfi)$, where $\bfi$
represents the measurement record.  Obviously, $\bfi$ refers to
different events for ideal and realistic detection: it is a record
of the times of photodetections and of avalanches, respectively.
The initial distribution can be found from \erf{drive} with $x$
taken to have a flat distribution.  The result is \beq
P_{0}(\Omega)=\frac{1}{\pi\sqrt{\Omega_{{\rm
max}}^{2}-\Omega^{2}}}. \label{InitProb} \eeq The unconditioned
evolution of the atom (which is independent of the scheme, or
quality, of detection) is given by the master equation (ME) \beq
\dot\rho(t)=-\frac{i\Omega}{2}[\hat\sigma_{x},\rho(t)]
+\Gamma{\cal D}[\hat\sigma]\rho(t)={\cal L}_{\Omega}\rho(t).
\label{master}\label{ME} \eeq Here $\Omega$ is assumed known,
$\Gamma$ is the spontaneous emission rate (which we will set equal
to unity), $\hat\sigma$ is the atomic lowering operator, and
$\hat\sigma_{x} = \hat\sigma + \hat\sigma\dg$. The superoperator
${\cal D}$ gives the damping of the system into the environment
and is defined as \cite{WisMil93c} \beq {\cal
D}[\hat{a}]\rho=\hat{a}\rho \hat{a}\dg -\half\cu{\hat{a}\dg
\hat{a}\rho+ \rho \hat{a}\dg\hat{a}}. \eeq A comparison of ideal
and realistic detection is given in Fig.~\ref{StateEstTrajProb}.
In this figure $\Omega_{\rm true} = 4\Gamma$, but the observer
does not know this, but rather has knowledge described by the
prior distribution (\ref{InitProb}) with  $\Omega_{\rm
max}=10\Gamma$. As time goes on, the detection record reveals more
about $\Omega_{\rm true}$, and this probability is updated in a
Bayesian manner. Note that for both ideal and realistic detectors,
the distribution always remains symmetric. This is because the
sign of $\Omega$ cannot be determined from direct detection.
Photoemissions depend only upon the mean of $\hat\sigma_z$, and
that is independent of whether the TLA state rotates clockwise or
counterclockwise around the $\hat\sigma_x$ axis.

\begin{figure}
\includegraphics[width=0.45\textwidth]{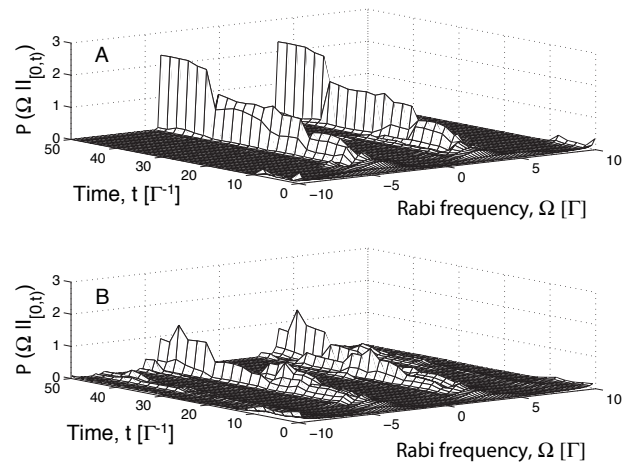}
\vspace{0.2cm} \caption{Plots (A) and (B) show how the probability
distribution for $\Omega$ typically evolves while continuous
direct detection is performed on the output of the quantum system
(a TLA inside an optical cavity). Plot (A) is for ideal detection,
while (B) is for realistic detection. The parameters used for the
photon counter  were $\eta=100\%$, and $\gamma_{{\rm r}}=7\Gamma$,
$\tau_{{\rm dd}}=0$, $\gamma_{{\rm dk}}=0$.  The value of $\Omega$
used to generate $\bfi$ was $\Omega_{{\rm true}}=4\Gamma$.  The
maximum allowed value for $\Omega$ is $\Omega_{{\rm
max}}=10\Gamma$. } \protect\label{StateEstTrajProb}
\end{figure}

In Fig.~\ref{StateEstTrajProb}, we have chosen a realistic
detector whose only imperfection is a finite bandwidth equal to 7$\Gamma$
(see figure caption). It can be seen that the
degradation of the quality of measurement is quite large even though our
`realistic' detector is well beyond current technology.
If we choose the realistic parameters for an APD (avalanche photodioide)
used in Refs.~\cite{WarWisMab02,WarWis03b} we obtain results that are
so far from those of ideal detection that a comparison lacks interest.

The fact that the chosen detector parameters give results that are still
poor compared to ideal measurement
is a little surprising.  Very few photodetections are missed under
realistic detection with a
finite $\gamma_{{\rm r}}$ being the only imperfection.  It
illustrates that the rate of information gain is very sensitive to
an uncertainty in the state of the TLA after a jump.

The plot for realistic detection, Fig.~\ref{StateEstTrajProb} (B),
also shows that it is difficult to distinguish between
$\Omega_{{\rm true}}$ and an integer multiple of $\Omega_{{\rm
true}}$, say $m\Omega_{{\rm true}}$ (for this figure this occurs
only for $|\Omega_{{\rm true}}|$ and $2|\Omega_{{\rm true}}|$). By
$\Omega_{{\rm true}}$ we mean the actual driving strength that
exists, but which is initially unknown to the observer. This
difficulty exists because the evolution for these two proposed
values of the driving strength
 will place the TLA in the excited state
at the same time every integer oscillation (at the smaller
frequency). Hence, for certain times there will be a strong
correlation between the probabilities of decay.  Obviously,  a
photoemission can still occur when a driving strength of
$2|\Omega_{{\rm true}}|$ would place the TLA very near ground
state, thus ruling out that possible value.  Whether these decays
that are discriminatory in this sense are detected or not depends
on the choice (and duration) of the quantum trajectory in
question. In Fig.~\ref{StateEstTrajProb} (B) a small but,
persistent (at least up to $t=50$) probability for
$|\Omega|=2|\Omega_{{\rm true}}|$ exists.

It is worth mentioning that the plots in
Fig.~\ref{StateEstTrajProb} are based on the `same'
photo-absorptions in the ideal and realistic cases.  This is
possible using the simulation technique of Refs.~\cite{WarWis03b},
in which the realistic quantum trajectory can be assured to be
consistent with the ideal quantum trajectory.

\section{Parameter and State Estimation Theory}

We now briefly present the theory of parameter and state estimation
in the context that is relevant to this paper.
The specific problem that we wish to address involves calculating the new
probability distribution for $\Omega$ and also obtaining the best estimate of
the TLA state, based on the measurement record.
Bayesian statistics are required, with the new distribution, based on $\bfi$,
given by \cite{BayesBook}
\beq
P(\Omega|\bfi)=\frac{P(\bfi|\Omega)
P_{0}(\Omega)}{\int P(\bfi|\Omega)
P_{0}(\Omega)d\Omega}.
\label{bayesPOm}
\eeq
The best estimate of the TLA state is found from \cite{Jay}
\beq
\rho_{{\bf I}}(t)=\frac{\int\tilde\rho_{{\bf I},\Omega}
(t) P_{0}(\Omega)d\Omega }
{\int P(\bfi|\Omega) P_{0}(\Omega)d\Omega},
\label{best2}
\eeq
where $\tilde\rho_{{\bf I},\Omega}(t)$ is the unnormalised state of
the TLA given that there was a measurement record $\bfi$ and that the
driving strength was $\Omega$.

In this paper we leave the state unnormalised even after jumps, so
that \beq \rho_{{\bf I},\Omega}(t)=\frac{\tilde{\rho}_{{\bf
I},\Omega}(t)}{P(\bfi|\Omega)} \eeq meaning that $P(\bfi|\Omega)$
is actually equal to the norm of the TLA state, when it is evolved
with $\Omega$ under $\bfi$. Ideally $\bfi$ would be an
experimental record, but for convenience we simulate it using
$\omtrue$ and then `forgetting' it.  The integrals in
\erfs{bayesPOm}{best2} require the simulation of the evolution for
the entire range of $\Omega$.  This means that the process is very
time consuming as many trajectories (for the different $\Omega$'s)
have to be run in order to obtain a single trajectory for
$P(\Omega|\bfi)$ or $\rho_{{\bf I}}(t)$.

In order to treat realistic detection, it is necessary  to
describe not just the state of the TLA, but also the state of a
detector  (treated classically) which is correlated with the
atomic state \cite{WarWisMab02,WarWis03a}. We do this by
introducing a set of three operators $\rho_{{\bf I},s}(t)$ for
this supersystem, where here $s=0,1,2$ represents the three
classical detector states. These correspond to the ready state
($s=0$), the avalanching state ($s=1$) and the resetting state
($s=2$). Refer to Fig.~\ref{Model} for details. The state of the
TLA is defined as $\rho_{\bf I}(t) = \rho_{{\bf I},
0}(t)+\rho_{{\bf I}, 1}(t) + \rho_{{\bf I}, 2}(t)$. Equation
(\ref{best2}) is trivially generalized to give the best estimate
of the supersystem state $\rho_{{\bf I},s}(t)$: \beq \rho_{{\bf
I},s}(t)=\frac{\int\tilde\rho_{{\bf I},\Omega,s} (t)
P_{0}(\Omega)d\Omega } {\int P(\bfi|\Omega) P_{0}(\Omega)d\Omega}.
\label{best2SuperSys} \eeq

The three equations, \erfs{bayesPOm}{best2} and \erf{best2SuperSys},
are unusable in their present form as $P(\bfi|\Omega)$ and the
norm of $\tilde\rho_{{\bf I},\Omega}(t)$ (which is equal to $
P(\bfi|\Omega)$) will become extremely small
as $t$ increases.  This will be true even for $\Omega=\omtrue$, as
can be seen by considering the probability of a string of measurement
results occurring.  The probability of a jump in a given infinitesimal interval is never greater than $\Gamma dt$, so that the norm of the state after $m$ jumps would be less
than  $(\Gamma dt)^{m}$ (since the no-jump evolution also decreases the norm of the state).
 In simulations, a time step of
$dt=10^{-4}\Gamma^{-1}$ is typical.  Thus, numerical error introduced
by computer round-off necessitates a different approach which we now explain.

\subsection{Linear Quantum Trajectories}

In linear quantum trajectories \cite{Wis96a}, the state is
multiplied by a predetermined factor after every measurement
event.  The normalisation constants are chosen so that the norm of
the states corresponding to the most likely values of $\Omega$
stay relatively close to unity.  Each possible measurement result,
$r$, has an `ostensible'  \cite{Wis96a} probability, $\Lambda (r)$
associated with it, that we use as the normalising factor.
Therefore, we have \beq \bar{\rho}_{{\bf
I},\Omega,s}(t)=\frac{\tilde{\rho}_{{\bf I},\Omega,s}(t)}{\Lambda
(\bfi)}, \label{Normalise} \eeq where $\bar\rho_{{\bf
I},\Omega,s}$ is the supersystem state when linear quantum
trajectories are used and $\Lambda (\bfi)$ is the ostensible
probability for getting $\bfi$, \beq \Lambda (\bfi)=\Lambda
(r_{k})\Lambda (r_{k-1})\ldots\Lambda (r_{1}). \eeq Here, we have
assumed that the time interval $[0,t)$ has been divided into a
very large number, $k = t/dt$, of discrete measurement times. The
actual probability of getting $\bfi$, given $\Omega$ is \beq
P(\bfi|\Omega)=\Lambda (\bfi)\sum_{s}{\rm Tr}[\bar{\rho}_{{\bf
I},\Omega ,s}(t)], \eeq where the trace is over the TLA and the
summation is over the detector states.

This allows \erf{bayesPOm} and \erf{best2SuperSys} to be written
as \cite{Jay} \beq P(\Omega|\bfi)=\frac{\sum_{s}{\rm
Tr}[\bar{\rho}_{{\bf I},\Omega,s} (t)]P_{0}(\Omega)}{\int
\sum_{s}{\rm Tr}[\bar{\rho}_{{\bf I},\Omega,s}(t)]P_{0}
(\Omega)d\Omega} \label{bayesPOm2} \eeq and \beq \rho_{{\bf
I},s}(t)= \frac{\int\bar\rho_{{\bf I},\Omega,s}
(t)P_{0}(\Omega)d\Omega } {\int \sum_{s} {\rm Tr}[\bar\rho_{{\bf
I},\Omega,s} (t)]P_{0}(\Omega)d\Omega}. \label{best3} \eeq Since
${\rm Tr}[\bar\rho_{{\bf I},\Omega,s}(t)]$ is of order unity for
the most likely $\Omega$'s the problems associated with using
${\rm Tr}[\tilde\rho_{{\bf I},\Omega,s}(t)]$ have been avoided.
The best estimate of the system state allows the conditional
evolution of a single stochastic record to be investigated, while
$P(\Omega|\bfi)$ will allow calculation of the information gain
(see following).

In this paper, we choose the ostensible probabilities to be \beq
\Lambda(1) = \epsilon dt, \eeq where $r=1$ indicates the detection
of a photon (ideal detector) or an avalanche (realistic detector).
No detection ($r=0$) has an ostensible probability
$\Lambda(0)=1-\epsilon dt$. With these ostensible probabilities,
the linear quantum trajectory equations for the supersystem
describing realistic detection \cite{WarWisMab02,WarWis03a} become
\bqa d\bar{\rho}_{0}(t)&=&dt\left({\cal L}_\Omega-\gamma_{{\rm
dk}}-\eta\Gamma{\cal
J}[\hat\sigma]+\epsilon\right)\bar{\rho}_{0}(t) \nl-d{\cal
N}(t)\bar{\rho}_{0}(t) {+}d{\cal N}(t)(t-\tau_{{\rm dd}})
\bar{\rho}_{2}(t)
\label{dp0epsilon}\\
d\bar{\rho}_{1}(t)&=&dt\left({\cal L}_\Omega-\gamma_{{\rm
r}}+\epsilon\right)\bar{\rho}_{1}(t) +dt\left(\eta\Gamma{\cal
J}[\hat\sigma] +\gamma_{{\rm
dk}}\right)\nl\times\bar{\rho}_{0}(t){-}d{\cal N}(t)\bar{\rho}_{1}
\label{dp1epsilon}\\
d\bar{\rho}_{2}(t)&=&dt\left({\cal
L}_\Omega+\epsilon\right)\bar{\rho}_{2}(t)+{d{\cal N}(t)}
\bar{\rho}_{1}(t)/{\epsilon}\nl{-}d{\cal N}(t)(t-\tau_{{\rm
dd}})\bar{\rho}_{2}(t). \label{dp2epsilon} \eqa Here $d{\cal
N}(t)=0,1$, the increment in the number of avalanches,
 is equivalent to the measurement result $r$ in the time interval $[t,t+dt)$.
We maintain the symbol $\bfi$ for the record of avalanches in the interval $[0,t)$, but
 for clarity we have omitted the subscripts ${\bf I}$ and $\Omega$ for the $\rho(t)$'s.
 The superoperator ${\cal J}$ is defined by \cite{WisMil93c}
 \beq
 {\cal J}[\hat{a}]\rho \equiv \hat{a}\rho \hat{a}\dg.
 \eeq

 For simulation purposes we take \beq \label{FluxDirect} \epsilon
= \eta  \frac{\Gamma\Omega_{\rm true}^{2}}{2\Omega_{\rm
true}^{2}+\Gamma^{2}}, \eeq the expected detection rate in steady
state (ignoring any dead time) when $\Omega$ is known. This is
allowed as the results of the simulation are independent of
$\epsilon$, thus we are free to choose any value.  Note that
renormalisation of $\rho(t)$ is occurring even when the detector
is resetting and the probability of an avalanche is strictly zero.
This illustrates the flexibility of the ostensible probabilities,
where the only concern is maintaining a norm of order unity.

\section{Conditional State Evolution}

Now that we have explained how the numerical results of Sec.~II
were obtained, we return to these simulations to investigate the
conditional state evolution in a typical realistic quantum
trajectory. In Fig.~\ref{StateEstTrajZ}  we see how $z(t)$ (the
mean of $\hat\sigma_z$) behaves under a variety of circumstances.
Specifically, we consider known $|\Omega_{{\rm true}}|$ and ideal
detection (thin dashed line), unknown $|\Omega_{{\rm true}}|$
(that is, an initial distribution given by \erf{InitProb}) and
ideal detection (thin solid line), known $|\omtrue|$ and realistic
detection (thick dashed line) and, finally, unknown $|\Omega_{{\rm
true}}|$ and realistic detection (thick solid line). The reason
why a comparison is made between the trajectories for a TLA
randomly placed in the standing wave of the field and trajectories
of known $|\omtrue|$, but unknown sign, is that at $t=\infty$ the
sign of $\omtrue$ will still not be known, as discussed in
Sec.~\ref{condprob}.  This allows us to reveal how far the
initially unknown $|\omtrue|$ trajectories are from their optimal
form, given the constraints of direct detection.

\begin{figure}
\includegraphics[width=0.45\textwidth]{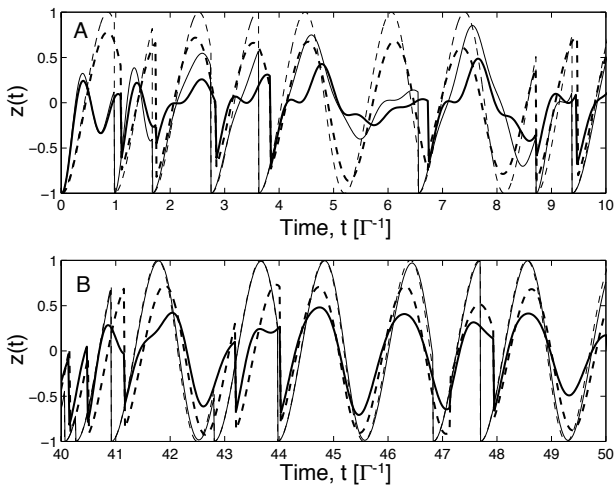}
\vspace{0.2cm} \protect\caption{Two segments of evolution of
$z(t)$ in a typical trajectory are shown.  The legend is as
follows: thin-lines are for ideal detection, thick-lines are for
realistic detection, dashed is for known $|\Omega_{{\rm true}}|$
and solid is for unknown $|\Omega_{{\rm true}}|$. Thus, to be
explicit, the thin dashed line is for $|\Omega_{{\rm true}}|$
known and ideal detection, the thin solid line is for
$|\Omega_{{\rm true}}|$ unknown and ideal detection, the thick
dashed line is for $|\Omega_{{\rm true}}|$ known and realistic
detection and, finally, the thick solid line is for $|\Omega_{{\rm
true}}|$ unknown and realistic detection. The detector and system
parameters (and time units) are as for
Fig.~\ref{StateEstTrajProb}.} \protect\label{StateEstTrajZ}
\end{figure}
%\clearpage

The conditioned state corresponding to a known $|\omtrue|$ is
formed from \beq \rho_{{\rm I}}(t)=(\rho_{{\rm
I},{\omtrue}}(t)+\rho_{{\rm I},{-\omtrue}}(t))/2. \eeq That is, it
is an equal mixture of the two conditioned states obtained with a
driving strength of $+\omtrue$ and $-\omtrue$.  Given that we take
the TLA as starting in the ground state, these two conditioned
states will have the same $z(t)$, but opposite sign for $y(t)$
resulting in $y(t)=0$ for the mixture as initially $P_{0}(\Omega)$
is an even mix of positive and negative $\Omega$. If the initial
state of the TLA (which is presumed to be known) gives a finite
value for $y(t=0)$ then the time of the first jump will give a
very small amount of information about $\omtrue$.  This is because
an early jump would tend to support the conclusion that the TLA
had been rotated on the Bloch sphere in the direction that takes
it towards the excited state first, while a later jump suggests
that the TLA was rotated in the opposite direction.  However, the
initial state is taken to be the ground state in our simulations.
Since for direction detection with no detuning $x(t)=0$
\cite{Car93b,WisMil93c}, we have \beq \rho_{{\rm
I}}(t)=(I+z(t)\hat\sigma_{z})/2, \eeq where $z(t)$ could be
obtained from either the $+\omtrue$ or $-\omtrue$ trajectory. It
should now be clear why only $z(t)$ is plotted.  In fact, the
purity can also be inferred from the $z(t)$ plot as $p_{\rm
I}(t)=(1+z^{2}(t))/2$.

As for Fig.~\ref{StateEstTrajProb}, the realistic trajectory is
forced to be consistent with the ideal trajectory. This is
evidenced by the jump times being correlated. Because of the
detail of these plots we show two portions of the evolution only,
illustrating how improved knowledge over time of the dynamical
parameter, $\Omega$, allows closer adhesion to the maximum
knowledge trajectory (thin dashed line for ideal detection and
thick dashed line for realistic detection).

In the first $10\Gamma^{-1}$ units of time the frequency of
oscillation of $z(t)$ for the realistic detection state (solid
black line) is poorly defined.  One can also see that the
frequency is faster than the true frequency, due to
$P_{0}(\Omega)$ being peaked at $\pm\Omega_{{\rm max}}$
\cite{Jay}. The small amplitude of oscillation is also due to the
lack of knowledge about $\Omega$.  The range of possible
frequencies ensures that when one value of $\Omega$ would place
the TLA in the excited state, another would place it near the
ground state, thus averaging the amplitude of oscillation of
$z(t)$ to below unity.

For ideal detection (thin solid line), the frequency of
oscillation is also shifting about, but the amplitude of the
oscillations have already noticeably increased in size by
$t=10\Gamma^{-1}$. Another feature is that jumps take $z(t)$ to
$-1$ (a pure state) for ideal detection even if $|\omtrue|$ is not
known.  This is because there is no time delay in conveying the
decay of the TLA to the observer. This is not true for realistic
detection where the jump is slightly delayed and the TLA will have
rotated out of the ground state by some unknown amount.

After $40\Gamma^{-1}$ time units, the ideal detection observer has
gained enough information about $|\Omega_{{\rm true}}|$ that
$z(t)$ matches well with the `ideal' $z(t)$
(Fig.~\ref{StateEstTrajZ} (B)). The realistic observer's $z(t)$ is
much closer to the ideal than at the start of the trajectory, but
some uncertainty still obviously exists.

\section{Information Gain}

We now move away from conditional dynamics and look at how the ensemble
averaged gain of information concerning $\Omega$ is affected by
realistic detection.
As in \cite{Jay}, the measure used to quantify the quality of information
gained about $\Omega$ is $\Delta I_{{\bf I}}$.  It is given, in bits, by
\bqa \Delta I_{{\bf I}} &=& \int d\Omega P(\Omega|\bfi) \log_{2}
P(\Omega|\bfi)
 \nl{-}\int d\Omega P_{0}(\Omega) \log_{2}
P_{0}(\Omega).
\label{bits}
\eqa
As the distribution for $\Omega$ becomes more sharply peaked
around $|\omtrue|$, $\Delta I_{{\bf I}}$ will increase.
Due to the
stochastic nature of $\bfi$, an ensemble of trajectories are run in
order to obtain the average, $\Delta I$, of $\Delta I_{{\bf I}}$ that
is characteristic of the measurement scheme.  Each member of the ensemble
is
formed by picking an $\omtrue$ randomly (but according to the
distribution \erf{InitProb}) and simulating the evolution based on a
stochastic trajectory corresponding to that $\omtrue$.  A single
trajectory for each member of an ensemble of $\omtrue$'s is
sufficient as the averaging of the stochasticity is done over the
multitude of $\omtrue$'s.

%figure showing distribution - could use to show reader how cannot
%determine sign of omega.  could have as a subplot the ideal
%distribution
%figure showing ensemble averages of purity, variance, delta I
%ideal, for number of omega
%figure showing ensemble averages of purity, variance, delta I
%realistic, for number of omega
%figure showing x,y,z,purity for ideal known, ideal unknown,
%realistic unknown
%after results describe how the simulation/statistics were calculated

A major difference between parameter estimation for ideal direct
detection and realistic direct detection is found when $\Omega_{{\rm
true}}\gg\Gamma$.  We investigate this regime by varying the value
of $\Omega_{{\rm max}}$ used in the simulations.  This works as the
most likely values for $\omtrue$ come from close to $\Omega_{{\rm
max}}$.  By increasing $\Omega_{{\rm max}}$ we are effectively
increasing the values of $\omtrue$ used.  This method corresponds
physically to the TLA being placed randomly in standing waves of
different amplitude.

In Fig.~\ref{Info} (A) the information gain, $\Delta I$, averaged
over an ensemble size of $1000$ (that is, $1000$ different
$\Omega_{{\rm true}}$'s), is shown for ideal detection.  The three
plots are for $\Omega_{{\rm max}}=5\Gamma,10\Gamma,20\Gamma$.  In
Fig.~\ref{Info} (B) the results are shown for the same values of
$\Omega_{{\rm max}}$ when the monitoring is done with realistic
detection. Error bars are included as the ensemble size in this
case is only $100$, due to the greater computational intensity. As
$\Omega_{{\rm max}}$ doubles, the information gain curves for
ideal detection are separated by approximately $1$ bit at large
$t$.  This does not indicate that the detection scheme is more
effective at large $\Omega_{{\rm max}}$, rather, it is an artifact
of the initial information content of the distribution
$P_{0}(\Omega)$.  This can easily be shown analytically. Consider
the integral \beq \int^{\Omega_{{\rm max}}}_{-\Omega_{{\rm max}}}
d\Omega \frac{1}{\pi\sqrt{\Omega_{{\rm max}}^{2}-\Omega^{2}}}
\log_{2}\frac{1}{\pi\sqrt{\Omega_{{\rm max}}^{2}-\Omega^{2}}}.
\eeq If the change $\Omega'_{{\rm max}}\rightarrow m\Omega_{{\rm
max}}$ is made, a change of integration variable to
$\Omega'=m\Omega$ allows the information of the new distribution
to be expressed in terms of the old one.  The result is \beq
\Delta I_{{\bf I},(m\Omega_{{\rm max}})}=\Delta I_{{\bf
I},(\Omega_{{\rm max}})} -\log_{2}m. \eeq Thus, every time
$\Omega_{{\rm max}}$ is doubled, the initial information decreases
by $1$ bit.  Since the initial information is being subtracted
from the information at time $t$, we can conclude that the width
of the $P(\Omega)$ distributions, for different $\Omega_{{\rm
max}}$, in the long time limit are roughly equal for ideal
detection.

\begin{figure}
\includegraphics[width=0.45\textwidth]{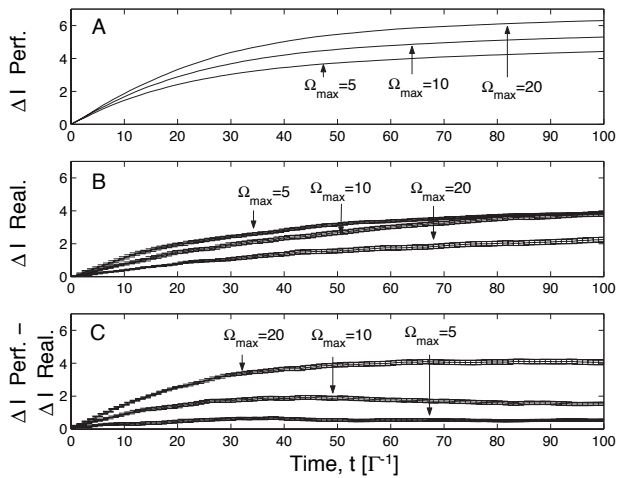}
\vspace{0.2cm} \caption{In plot (A) the ensemble averaged
information gain for ideal detection is shown for three different
values of $\Omega_{{\rm max}}$. In plot (B) the ensemble averaged
(over $100$ trajectories) information gain for ideal detection is
shown for the same values of $\Omega_{{\rm max}}$.  In plot (C)
the difference between ideal and realistic detection is shown. }
\protect\label{Info}
\end{figure}

This is qualitatively different from the case of realistic
detection where the information gain for $\Omega_{{\rm
max}}=5\Gamma$ and $10\Gamma$ are roughly the same and
$\ommax=20\Gamma$ sees a significant decrease.  Taking into
account the fact that as $\ommax$ increases less information is
being subtracted away from the information at time $t$, it is
clear that realistic detection is becoming considerably worse at
distinguishing between candidates for $\omtrue$.  This is more
clearly illustrated in Fig.~\ref{Info} (C) which shows the
difference between ideal and realistic detection information gain
for the three values of $\ommax$.  We now try to understand these
trends by discussing what features of a measurement record allow
the information to be accrued.

The most simple (and naive) way to determine the actual driving
strength is to analyse the mean steady-state photon counting rate
$\langle {\bf I} \rangle_{\rm ss}$. For ideal direct detection
this is
\begin{equation}\label{Iss}
  \langle {\bf I}
  \rangle_{\rm ss}=\frac{\Gamma\Omega^2}{2\Omega^2+\Gamma^2}.
\end{equation}
 By using \erf{Iss} the mean flux can be matched
with a particular $\Omega$, given that $\Gamma$ is known. However,
the flux asymptotes to $\Gamma/2$ at large $\Omega$, essentially
making one value of $\Omega$ indistinguishable from another.
Because ideal detection still gives information at large
$\Omega_{{\rm max}}$ we deduce that the specific times at which
detections are occurring also play a role in determining
$\Omega_{{\rm true}}$.

To consider this effect in more detail we look at the waiting time
distribution for photon detections. In perfect direct detection
\cite{wtime} \beq
w(\tau)=\frac{\Omega^{2}}{\Omega^{2}-(\Gamma/2)^{2}}e^{-\Gamma\tau/2}
\sin^{2}\{\half[\Omega^{2}-(\Gamma/2)^{2}]^{1/2}\tau\},
\label{wTimeErf} \eeq where $w(\tau)$ is the probability density
of a photon being emitted by the TLA at a time $\tau$ after the
previous emission. The waiting time, $w(\tau)$, is zero at
$\tau=0$ as the TLA in the  ground state after a detection. A plot
of this function for $\Omega=5\Gamma$, with $\Gamma=1$ is given in
Fig.~\ref{wtimeFig} (A).  The point that is meant to be taken from
this graph is that there are oscillations that make certain
emission times more likely (when the TLA is in the excited state)
and other times prohibited (when it is in the ground state), given
a certain driving strength.  These oscillations, which take
$w(\tau)$ to zero, are present for all $\Omega>\Gamma/2$, which
includes the regime of interest here. This allows the narrowing of
the distribution for $\Omega$, based on a measurement record,
despite the fact that very little information can be obtained from
the mean flux rate.

The waiting time distribution for realistic detection is much more
complex.  In this case, $\tau$ would be the waiting time between
avalanches of the photodiode, rather than emissions of the TLA.
One of the difficulties lies in the fact that there is not a
definite state of the atom after an avalanche, due to the finite
response time. Thus, initial conditions for the TLA must be chosen
as being characteristic of the post-avalanche state.  We chose
them to be the solution to the ME (\ref{ME}) averaged over the
time for the avalanche to mature.  That is, \beq \rho_{{\rm
aval.}}=\gamma_{{\rm r}}\int_{0}^{\infty}e^{-\gamma_{{\rm
r}}t}\rho(t)dt, \label{AvaICs} \eeq where $\rho(t)$ is the
solution to the ME given that at $t=0$ the TLA is in the ground
state.  Of course, this will only be true if the dark counts are
ignored, which we will do.  The integral in \erf{AvaICs} can be
carried out analytically with $x=0$ and the $y$ and $z$ components
of the state being \bqa y_{{\rm
aval.}}&=&\frac{2\Omega(1+\gamma_{{\rm r}})}
{(1+3\gamma_{{\rm r}}+2\gamma_{{\rm r}}^{2}+2\Omega^{2})},\label{yAval}\\
z_{{\rm aval.}}&=&-\frac{(2\gamma_{{\rm r}}+1)(1+\gamma_{{\rm r}})}
{(1+3\gamma_{{\rm r}}+2\gamma_{{\rm r}}^{2}+2\Omega^{2})},
\label{zAval}
\eqa
where we have set $\Gamma=1$.

If we take the simplest case where  $\tau_{dd}=0$ and $\gamma_{dk}=0$, then
we can use
these  as the initial conditions for the ready state of the
detector. (The case where $\tau_{dd}\neq0$ is only a little more complicated,
because the solutions in \erfs{yAval}{zAval} can be evolved
forward a time $\tau_{{\rm dd}}$ with the ME to give the state of
the TLA when the detector becomes ready to detect photons once
more.) A series of $6$ linear coupled differential equations must
be solved in order to determine $\gamma_{{\rm r}}{\rm
Tr}[\tilde{\rho}_{1}(\tau)]=w(\tau)_{{\rm aval.}}$, which is equal
to the probability of an avalanche at time $\tau$ (the waiting
time). Here, $\tilde{\rho}_{1}(\tau)$ is the solution to the
unnormalised evolution equations given in Eqs.~(\ref{dp0epsilon})
and (\ref{dp1epsilon}) with $\epsilon=0$ and $d{\cal N}\equiv 0$
(no avalanches) from $\tau_{\rm dd}$ to $\tau$.

The $6$ real variables that are necessary can be expressed as a vector
$\vec{r} = (\tilde{y}_{0},
\tilde{z}_{0}, \tilde{P}_{0}, \tilde{y}_{1},
\tilde{z}_{1}, and \tilde{P}_{1})^{\rm T}$, where the the subscripts
refer to the detector state and the $P$'s are the probability of
occupation.  The equations of motion for this vector is
$
d\vec{r}/dt = {\bf A}{\vec r}
$
where ${\bf A}$ is the following matrix
\beq
\left(\begin{array}{cccccc}
-\half & 0 & -\Omega & 0 & 0 & 0\\
0 & -\half-\gamr & 0 &-\Omega & 0 & 0 \\
\Omega & 0 & -1+\frac{\eta}{2} & 0 & -1+\frac{\eta}{2} & 0  \\
0 & \Omega & -\frac{\eta}{2} & -1-\gamr & -\frac{\eta}{2} & -1\\
0&0& -\frac{\eta}{2} &0 & -\frac{\eta}{2} & 0\\
0&0& \frac{\eta}{2} &0 & \frac{\eta}{2} & -\gamr
\end{array}\right),
\eeq
%\begin{widetext}
%\bqa
%\frac{d}{d\tau}\left(\begin{array}{ll}
%\tilde{y}_{0}\\
%\tilde{z}_{0}\\
%\tilde{y}_{1}\\
%\tilde{z}_{1}\\
%\tilde{P}_{0}\\
%\tilde{P}_{1}
%\end{array}\right)=
%\left(\begin{array}{cccccc}
%-\half & 0 & -\Omega & 0 & 0 & 0\\
%0 & -\half-\gamr & 0 &-\Omega & 0 & 0 \\
%\Omega & 0 & -1+\frac{\eta}{2} & 0 & -1+\frac{\eta}{2} & 0  \\
%0 & \Omega & -\frac{\eta}{2} & -1-\gamr & -\frac{\eta}{2} & -1\\
%0&0& -\frac{\eta}{2} &0 & -\frac{\eta}{2} & 0\\
%0&0& \frac{\eta}{2} &0 & \frac{\eta}{2} & -\gamr
%\end{array}\right)
%\left(\begin{array}{ll}
%\tilde{y}_{0}\\
%\tilde{z}_{0}\\
%\tilde{y}_{1}\\
%\tilde{z}_{1}\\
%\tilde{P}_{0}\\
%\tilde{P}_{1}
%\end{array}\right).
%\eqa
%\end{widetext}
 whereas before we are assuming $\gamma_{{\rm dk}}=0$ and
$\Gamma=1$.
The variable of interest is then $w(\tau)_{\rm aval}=\gamma_{{\rm
r}}\tilde{P}_{1}(\tau)$

\begin{figure}
\includegraphics[width=0.45\textwidth]{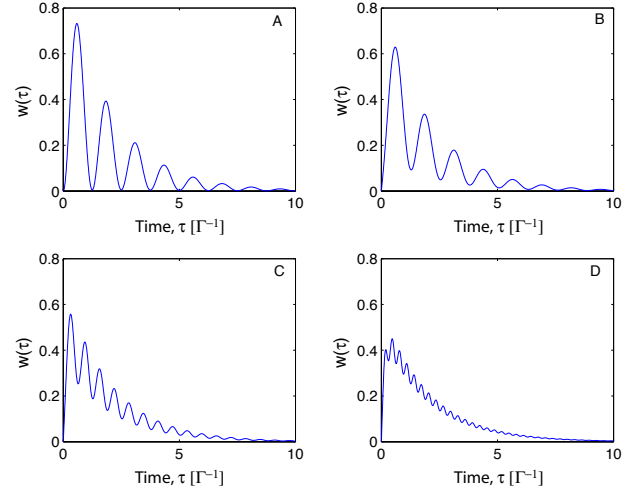}
\vspace{0.2cm} \caption{In plot (A) the waiting time distribution
for ideal detection is shown for $\Omega=5\Gamma$.  In plots (B),
(C) and (D) the waiting time between avalanches for realistic
detection is shown for the values of $\Omega=5\Gamma$,
$\Omega=10\Gamma$ and $\Omega=20\Gamma$.  The photon counter
parameters are as for Fig.~\ref{StateEstTrajProb}.}
\protect\label{wtimeFig}
\end{figure}
%\clearpage

Plots of $w(\tau)_{{\rm aval.}}$ with $\Omega=5\Gamma$,
$10\Gamma$,  and $20\Gamma$ are given in Fig.~\ref{wtimeFig} (B),
(C) and (D).  Most notable is that although there are
oscillations, they do not take $w_{{\rm aval.}}(\tau)$ to zero.
The random response time of the photodetector is washing out the
peaks and troughs of the ideal detection waiting time
distribution, making it much more difficult to distinguish between
$\Omega$'s on the basis of the specific times of the avalanches.
This effect becomes more pronounced as $\Omega$ increases.  The
relevant characteristic time scale is the response time of the
detector. For $\Omega\agt\gamr$, the TLA may have rotated out of
the ground state significantly by the time the avalanche occurs.
If the state of the TLA is smeared out and information contained
in the higher order moments of the waiting distribution is lost,
then there is no way to determine $|\omtrue|$ when it is large.
This is the reason for the drop-off of the information gain for
realistic detection that is seen in Fig.~\ref{Info}.

Although we have done our simulations with a dead time
$\tau_{dd}=0$, the finite response bandwidth leads to an effective
dead time of $\gamma_{r}^{-1}$, which leads to an effective drop
in the efficiency of the detector. To rule inefficiency as the
cause of the loss of structure in $w(\tau)$ as $\Omega$ increases
we consider the waiting time distribution for a detector of
efficiency $\eta$. Using the formula given in Ref.~\cite{wtime}
%as \bqa
%w(\tau)&=&\sum_{i=1}^{3}
%\Big{[}\frac{\eta\Gamma\Omega^2(s-s_i)}{2(s_i-s_1)(s_i-s_2)(s_i-s_3)}\Big{]}\Big{|}_{s=s_i}\nl\times
%\exp(s_i \tau), \label{wTimeEff} \eqa where $s_1$, $s_2$, and
%$s_3$ are the roots of the cubic
%\begin{equation}\label{cubic}
%  s(s+\Gamma)(s+\Gamma/2)+\Omega^2(s+\eta\Gamma/2)=0.
%\end{equation} To illustrate the affect of this waiting time, the
%roots of this equation where calculated numerically, and plots of
we have calculated
$w(\tau)$ with $\Omega=5\Gamma$ , $10\Gamma$, and $20\Gamma$.
Plots are
given in Fig.~\ref{wtimeFigeff} (B), (C) and (D) (with
$\eta=70\%$) while Fig.~\ref{wtimeFigeff} (A) is a reproduction of
Fig.~\ref{wtimeFig} (A) ($\Omega=5\Gamma$ and $\eta=100\%$). Here
it is observed that the smearing out of $w(\tau)$ (for detectors
with only an inefficiency) is due only to this inefficiency. That
is, it is independent of $\Omega$. Thus the smearing out the
waiting time distribution as a result of increasing $\Omega$
for a realistic detector is
brought about by the
finite bandwidth, $\gamma_r$.

\begin{figure}
\includegraphics[width=0.45\textwidth]{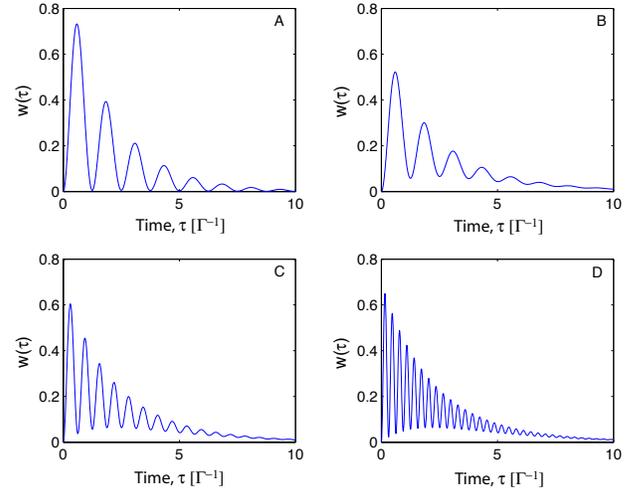}
\vspace{0.2cm} \caption{In plot (A) the waiting time is plotted
for $\Omega=5\Gamma$ and $\eta=100\%$.  In plots (B), (C), and (D)
the waiting time is plotted for $\Omega=5\Gamma$ ,
$\Omega=10\Gamma$ and $\Omega=20\Gamma$ and $\eta=70\%$.}
\protect\label{wtimeFigeff}
\end{figure}
%\clearpage

\section{Discussion}

We have investigated the effect of imperfections in realistic
detectors upon the problem of quantum state and parameter
estimation by continuous monitoring of an open quantum system.
Specifically, we have re-examined the system of a two-level atom
(TLA) with an unknown driving strength introduced in
Ref.~\cite{Jay}. Considering only direct photodetection, we find
that the most significant effect comes from a finite bandwidth of
the detector. This corresponds to a randomness in the response
time of the photodiode (the time from photon absorption to the
time when the avalanche reaches a pre-set threshold).

We find that unless the bandwidth is significantly greater than
the Rabi frequency, the observer's ability to obtain information
about the unknown Rabi frequency, and about the state of the atom,
is severely compromised. This is because the waiting time
distribution between avalanches is smeared out by the bandwidth,
losing the Rabi oscillations that characterize it for an ideal
detector (see Fig. \ref{wtimeFig}). This result shows that the
Bayesian update method in Ref.~\cite{Jay} implicitly made use of
the structure of the waiting time distribution, rather than simply
the mean detection rate. This implies that even for an ideal
detector this parameter estimation problem is extremely nonlinear,
and requires the full power of Bayesian probability theory. Our
results thus have implications for quantum control \cite{Dohe2000}
in the presence of unknown parameters even for ideal detectors.

A final comment on the question of the physicality of the scenario
that we have constructed needs to be made.  In order to measure
the emissions of an atom with an efficiency anywhere close to
unity it would be necessary to place the atom in a micro cavity,
with the cavity mode being heavily damped so that the atom can be
regarded as emitting
 predominantly into the cavity output beam \cite{RicCar88,TurThoKim95}.
However, in this paper we have considered a situation in which the
position of the TLA is random, but fixed, across one wavelength of
a standing wave field.  This has the ramification that the
effective decay rate is also position-dependent and hence also
uncertain, scaling like the driving strength squared. Thus, in
reality, there would be uncertainty in the two dynamical
parameters $\Omega$ and $\Gamma$.  In principle, with ideal
detection, both these parameters could be found (apart from the
sign of $\Omega$) given a long enough measurement record, as the
waiting time distribution for ideal detection will be different
for different positions of the TLA in the standing wave. It would
be interesting to see how realistic direct detection
affects the rate gain of information about $\Omega$ and $\Gamma$.

Another question which would be interesting to answer is, given
that homodyne detection of the $y$ quadrature in \cite{Jay} was
shown to provide the greatest gain in information, does this
result also apply when realistic detection affects are taken into
account. Lastly, an interesting idea would be to investigate
this scenario with a frequency-filtered measurement scheme, such as in
Ref.~\cite{WisToo99}. This would require extending
the measuring apparatus to include many cavities (one for each
frequency) with each cavity having its own realistic detector.
Either of these schemes would require far greater numerical resources than the
(not insignificant) resources used here.

\begin{acknowledgments}
This work was supported by the Australian Research Council.
\end{acknowledgments}

\end{document}